# Topographic Temperature: A Maximum-Entropy State Description of Running-In Surfaces

**Boris Brodmann**[1*] and **Matthias Eifler**[2]

[1] OptoSurf GmbH, Ettlingen, Germany

[2] IU Internationale Hochschule, Juri-Gagarin-Ring, Erfurt, Germany

* *Corresponding author: boris.brodmann@optosurf.de*

## Abstract

Surface topography governs tribological performance, yet conventional parameters describe either amplitude statistics or spectral content in isolation. We introduce a scale-dependent framework that represents surface height and directional gradient as conjugate coordinates of a structural phase space. The elastic reference energy derived from Persson's contact mechanics theory defines a metric that couples surface geometry to the elastic half-space response. A maximum-entropy formulation yields a canonical state density. In the Gaussian limit this formulation recovers Persson's spectral description exactly, showing that the power spectral density is a complete contact mechanical descriptor only under Gaussian statistics. The associated Lagrange multiplier defines a topographic temperature in the sense of Grmela's multiscale thermodynamics and embeds the areal subsystem within a scale-dependent boundary potential. The framework is validated experimentally using ground and honed AISI 52100 steel discs before and after running in. The ground surface contracts toward lower entropy and elastic energy, whereas the honed surface expands into previously unoccupied states. These opposite trajectories become transparent only in the coupled height-gradient representation and highlight the role of the principal directional gradient for scale-aware surface metrology.

## 1. Motivation

Surfaces present a particular challenge in many research disciplines. Wolfgang Pauli is often quoted as having remarked: "God made the bulk, but the devil was responsible for the surface." This quotation points to the fact that interfaces in real systems frequently carry the physically relevant complexity. A vivid illustration is provided by Gabriel's horn (Torricelli's trumpet): a geometric body can possess a finite volume while its surface area becomes infinite under idealized geometry [1]. In physics, Bekenstein's analysis of black holes underscores this special role even more fundamentally, by coupling thermodynamic state variables to an area (the event horizon) rather than to the volume [2].

In tribology, this special status is directly relevant to practical applications. Interfaces separate the base and counter bodies as well as intermediate and surrounding media, and they evolve over time under a load spectrum comprising mechanical, thermal, chemical, and physical influences. These changes manifest as dissipative processes involving material and power losses, ultimately leading to various forms of wear [3]. The underlying mechanisms can each be explained on their respective scales. However, their overall effect emerges only from their interaction across scales.

Tribological phenomena are processes of open thermodynamic systems far from equilibrium: only through the continuous input of mechanical work can dissipation, material transport, and irreversible microstructural changes proceed as a process. A suitable reference framework is therefore the non-equilibrium thermodynamics of open systems and dissipative structures [4].

This becomes particularly evident during the running-in phase, in which - under favorable process conditions - states of very low wear rates can be achieved [5]. Archard formulated a process model in which asperities plastically deform under load until a predominantly elastic state is reached [6]. Complementarily, Godet's "third-body" concept describes the formation of a third body composed of wear particles and reaction products, acting as a functional intermediate structure within the tribological system [7]. Nosonovsky and Bhushan frame these processes thermodynamically and interpret protective layers and pattern formation as self-organized secondary structures with positive entropy production [8,9]. Mortazavi et al. propose considering Shannon entropies derived from height distributions for specific aspects of surface evolution [10]. Figure 1b classifies the running-in process as a state evolution: Under continuous energy input, the tribosystem can transition from strongly dissipative initial regimes into a stationary operating state with lower loss measures [5].

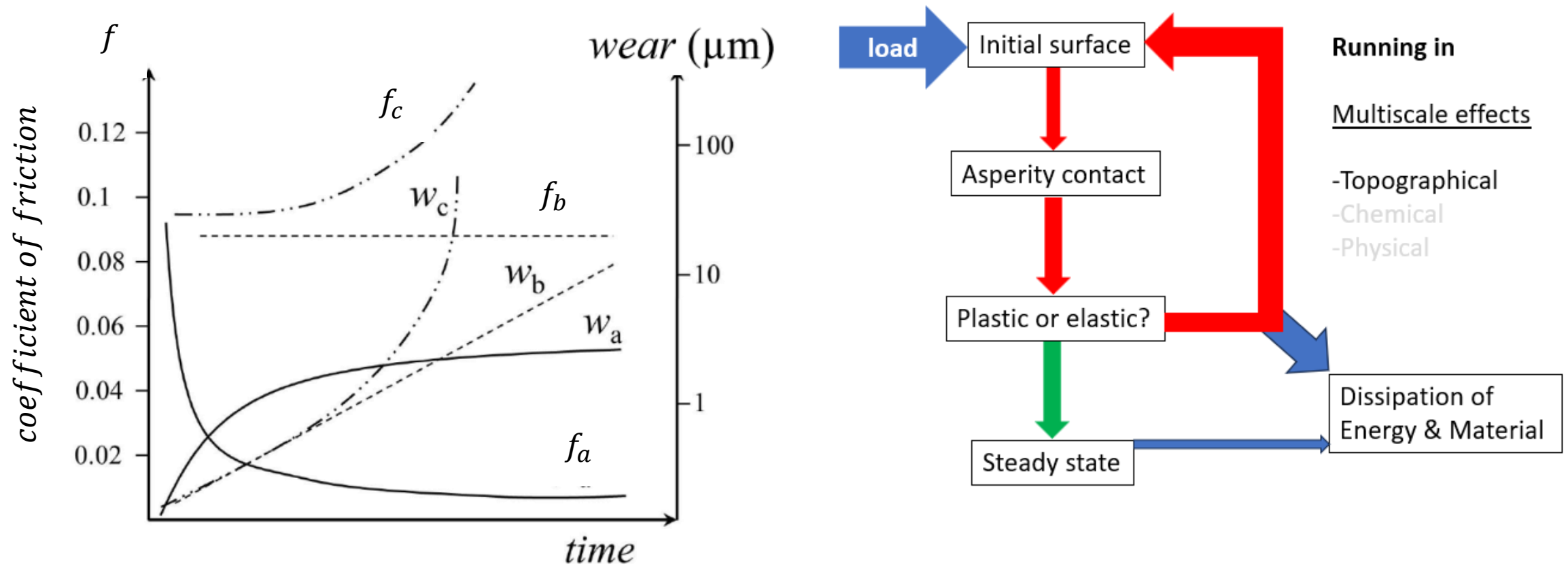

Fig. 1a) Different wear and friction behavior of tribological systems, adapted from [5, Scherge], Running-In Behavior of Tribological Systems, Lubricants 2018, 6, 54, CC BY 4.0

Fig. 1b) Process of topographical running in and the "exit" point with the steady state regime

The corresponding macroscopic signature is reflected in the temporal evolution of the friction coefficient and wear rate (Fig. 1a). The macroscopic evolution is determined by the local deformation and reaction regime in the load-bearing contact areas. At the same time, the physically relevant contact zone in many technical tribosystems is a moving, "buried" interface in which local stresses, temperatures, material transformations, and chemical reactions occur in a coupled manner. These quantities are usually accessible only indirectly in experiments, even though they dominate the macroscopic response (friction and wear). This leads to the necessity of describing the system state through suitable reduced state variables and modeling frameworks that structure the multiscale and multiphysics problem without requiring a complete microscopic representation [11].

The geometric contribution of surface properties is classically described using roughness or more general surface texture parameters. Greenwood and Williamson introduced the plasticity index as a contact-mechanical parameter that couples RMS roughness and mean asperity curvature, thereby providing a link between geometry and material response (elastic/plastic) [12]. For random roughness, these concepts were extended by stochastic surface models. Nayak characterizes roughness via the spectral moments of the power spectral density (PSD) [13]. Whitehouse and Archard discuss significant properties of random surfaces and simultaneously emphasize the influence of measurement and evaluation procedures on spectral quantities [14]. Persson highlights the central role of the PSD for the mechanics of rough contacts, incorporates self-affine scaling as a general descriptive element, and couples it in a scale-dependent manner to the elastic half-space [15]. The classification of topography as a stochastic process and the early use of fractal parameters were advanced, among others, by Sayles/Thomas and Brown et al. [16-18]. Also Aghababaei et al. demonstrated the fractal character of physical surfaces across large scale ranges [19]. At the same time, there is justified skepticism regarding the practical usability of fractal parameters, particularly due to the sensitivity of high-frequency components to measurement noise, limited resolution, and calibration [20,21]. In addition, Pawlus showed that many standardized surface texture parameters exhibit high correlation, as they are closely related and often derived solely from the stochastic distribution of height values [22].

Since friction and wear are not purely geometric phenomena but represent irreversible flows of energy and matter within a scarcely accessible contact zone, a functional description of topography must be embedded in a thermodynamic reference framework. This framework explicitly accounts for internal dissipation through frictional power as an independent process component within the mechanically affected zone (MAZ). Within this zone, topography, mechanics, chemistry, and mass transport are coupled in a scale-dependent manner, such that a single geometric parameter can only incompletely represent the system state [23]. Accordingly, height distributions and PSDs are not equivalent descriptions of a topography ensemble: for a given PSD, the height distribution typically does not follow uniquely and vice versa, with the exception of their shared coupling of amplitude via the second moment $R_q^2$, which corresponds to the zeroth spectral moment $m_0$ of the PSD [24]. A purely amplitude-statistical or purely spectral description therefore captures only partial aspects of the state of the topography.

The objective is to formulate the topography itself as a scale-dependent state space so that functional descriptors can be expressed as state variables of a consistent model. This aims to unify amplitude-statistical approaches and spectral contact models within a thermodynamic reference framework that conforms to the phenomenological state evolutions of tribological systems (running-in, stationary regimes, MAZ).

We therefore formulate these as an ensemble of a scale-dependent subsystem within an open thermodynamic overall system. To this end, we introduce a state description that couples the PSD-based elastic energy with the entropy of a shared state density and treats the ensemble as a subsystem of the mechanically affected zone with its own thermodynamic potential.

## 2. Thermodynamic Description of the MAZ and Scale-Dependent Surface States

Tribological systems can be regarded as open, dissipative many-body systems that exchange energy, momentum, and matter with their surroundings under external load. They typically operate far from equilibrium, since the contact geometry is locally heterogeneous and scale-continuously modified through plastification, film transfer, tribochemical processes, and changes in topography [23]. As a modeling framework, a coupling to effective reservoirs is therefore chosen. The mechanically affected zone (MAZ) [23] is described as an open subsystem of a coupled overall system that exchanges energy and mass with an effective bath at $T_{\text{bath}}$ and $\mu_{\text{bath}}$, while the external load, through the driving work flux $\dot{W}_{\text{mech}}$, drives the system into a non-equilibrium (stationary) state with positive entropy production.

Macroscopically, these processes manifest in measurable response quantities such as the friction coefficient $f(t, v)$ and the wear rate $w(t, v)$ (see Fig. 1a), whose explicit dependence on time and sliding velocity characterizes different operating states of the tribosystem. Figure 2 illustrates how mechanical power $\dot{W}_{\text{mech}}$ is supplied to the system via the normal force $F_N$ and tangential force $F_t$, while dissipation and mass transfer are released to the surroundings as fluxes $\dot{Q}_{\text{out}}$ and $\dot{N}_{\text{out}}$. The micro-geometry of the contacting bodies generates local pressure peaks and thus spatially concentrated activation within the MAZ.

Within the framework of Nonequilibrium Thermodynamics [25], the MAZ can be regarded as a hierarchical system in which individual scales are characterized by their own thermodynamic potentials $\Omega_i$. While a complete description would require coupling all relevant scales - from atomic dynamics and defect formation in the near-surface layer to mesoscale topography - the present work focuses on the topography as a scale-dependent subsystem within this multiscale structure.

For the thermodynamic modeling, the surface potential is first decomposed independently of any spectral resolution into two structurally distinct parts:

$$\Omega_{\mathrm{surf}}^{\mathrm{sub}} = \Omega_{\mathrm{bulk}}^{\mathrm{sub}} + \Omega_{\mathrm{area}}^{\mathrm{sub}}. \tag{1}$$

Here, $\Omega_{\mathrm{bulk}}^{\mathrm{sub}}$ denotes the potential of the near-surface volumetric part (the boundary layer with chemical, microstructural, and defect degrees of freedom). In the language of Multiscale Nonequilibrium Thermodynamics [25,26,29], this term represents the contribution of those microscopic details that cannot be expressed through the geometric state variable alone, it plays the role of the internal energy $\epsilon$ that Grmela and Öttinger introduce for the ignored degrees of freedom on the mesoscopic level. The second term, $\Omega_{\mathrm{area}}^{\mathrm{sub}}$, carries only the geometric boundary and is governed by the real surface area $A_{\mathrm{real}}$ as its central state variable.

For a surface represented as an areal graph $z(x, y)$, the real surface area relative to the projected reference area $A_0$ is given by

$$A_{real} = \int_{A_0} \sqrt{1 + \| \nabla z(x,y) \|^2} \, dA_0. \tag{2}$$

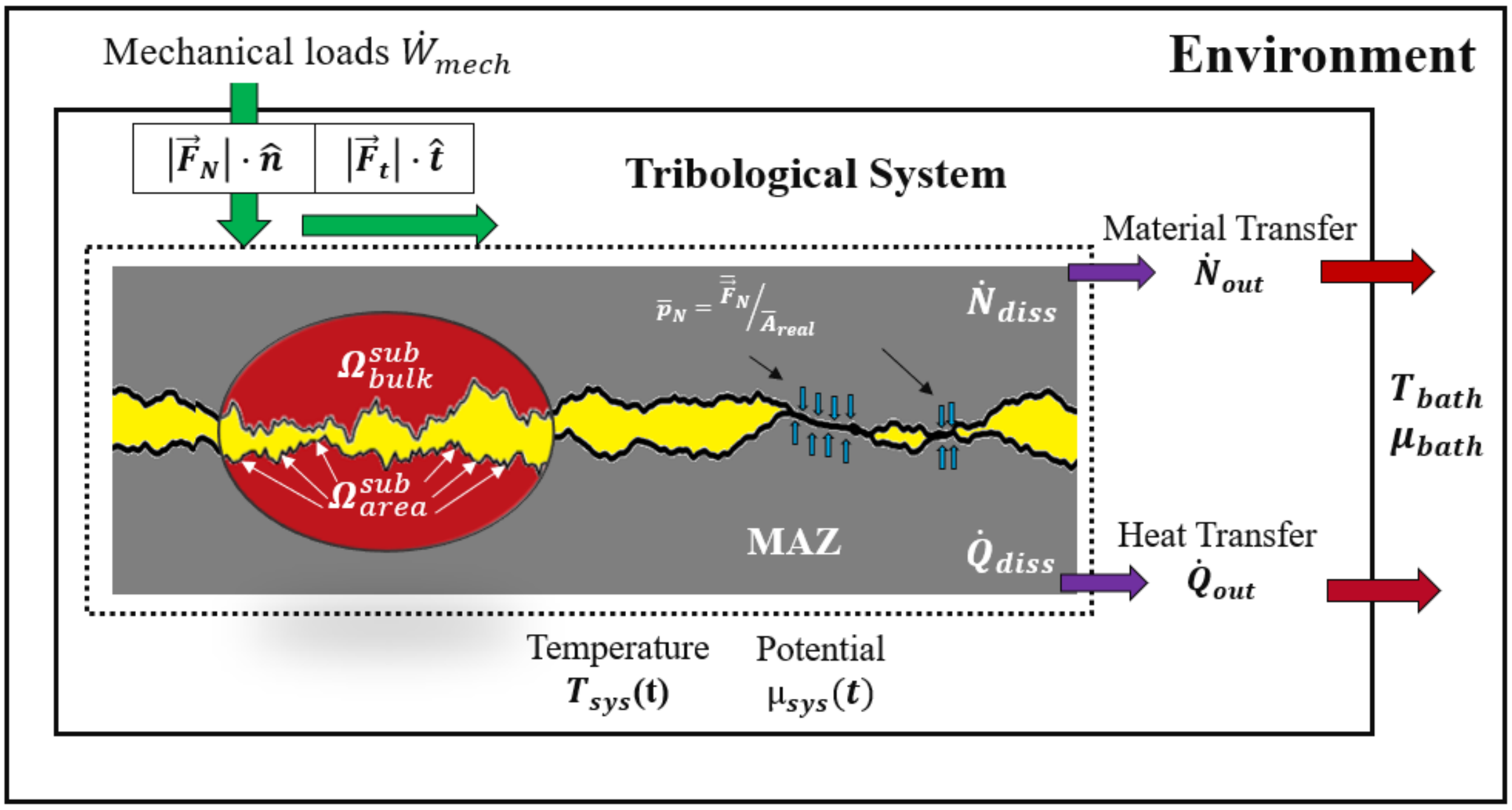

Fig. 2 Schematic model of the mechanically affected zone (MAZ) as an open, dissipative subsystem. Mechanical work $\dot{W}_{mech}$ drives the tribosystem, while heat and mass fluxes ($\dot{Q}_{out}$, $\dot{N}_{out}$) are transferred to an effective bath ($T_{bath}$, $\mu_{bath}$). Internal dissipation $\dot{Q}_{diss}$ and particle generation $\dot{N}_{diss}$ arise within the contact zone at system temperature $T_{sys}(t)$ and chemical potential $\mu_{sys}(t)$.

To treat the surface as a thermodynamic subsystem within the MAZ, we introduce a generalized thermodynamic potential as the appropriate state function. This potential represents the exchange of heat and matter through $T$ and $\mu_i$, and incorporates the geometry through a dedicated area term. Formally, this can be expressed as:

$$\mathrm{d}\Omega_{\mathrm{surf}}^{\mathrm{sub}} = -S\,\mathrm{d}T - \sum_i N_i\ \mathrm{d}\mu_i + \gamma\,\mathrm{d}A_{\mathrm{real}} \tag{3}$$

Here, $\gamma$ is the quantity conjugate to the real surface area,

$$\gamma = \left(\frac{\partial\,\Omega_{surf}^{sub}}{\partial\,A_{real}}\right)_{T,\{\mu_i\}}, \tag{4}$$

and denotes an effective surface energy per unit change in area. The chemical potential terms $\mu_i$ account for mass exchange between the surface and the bulk of the MAZ. In the subsequent analysis, however, these contributions are absorbed into $\Omega_{bulk}^{sub}$, so that the areal subsystem $\Omega_{area}^{sub}$ carries only the geometric degrees of freedom. The equation for $\mathrm{d}\Omega_{\mathrm{surf}}^{\mathrm{sub}}$ has the structure discussed by Landau and Lifshitz [27] for the equilibrium of a crystal with a free surface: the variation $\mathrm{d}\Omega_{\mathrm{area}}^{\mathrm{sub}} = \gamma\,\mathrm{d}A_{\mathrm{real}} = \min$ determines the equilibrium shape, and $\gamma$ corresponds to the surface tension $\alpha$ in Landau & Lifshitz, §159. In this limit, $A_{\mathrm{real}}$ is a smooth, Euclidean surface area, $\gamma$ is a material constant of the interface, and the two quantities form a proper conjugate pair.

Real engineering surfaces, however, are not smooth but exhibit roughness on many length scales. For self-affine surfaces, $A_{\mathrm{real}}$ diverges as the measurement resolution increases the surface area depending on the scale at which it is observed, and, because of the area divergence, the system is far away from its equilibrium. To make this dependence explicit, we introduce, in the sense of Persson [15] the dimensionless zoom parameter $\zeta$, which parametrizes the set of resolved roughness modes through a wavenumber cutoff:

$$k(\zeta) = \zeta\,k_0, \qquad k_0 := \frac{2\pi}{L}, \tag{5}$$

where $L$ is the lateral reference length (extent of the measurement field). Equivalently, this yields an effective lateral resolution of

$$\lambda(\zeta) = \frac{2\pi}{k(\zeta)} = \frac{\lambda_0}{\zeta}, \tag{6}$$

which, in surface-metrology terms, corresponds to a high-frequency band limitation (short-wavelength cutoff) [34,35]. The topography represented at zoom $\zeta$ is therefore understood as a band-limited field $z_\zeta(x,y)$ that contains only spectral components with $\parallel k \parallel \leq k(\zeta)$.

With this convention, the decomposition and the area acquire an explicit scale dependence:

$$\Omega^{sub}_{surf}(\zeta) = \Omega^{sub}_{bulk}(\zeta) + \Omega^{sub}_{area}(\zeta), \quad A_{real}(\zeta) = \int_{A_0} \sqrt{1 + \| \nabla z_\zeta(x,y) \|^2} \, dA_0. \tag{7}$$

Thus, $A_{\mathrm{real}}$ is no longer a fixed geometric quantity but a function of the resolved bandwidth. The conjugate variable $\gamma$ likewise becomes scale-dependent: $\gamma(\zeta)$.

For the spectral description, we treat the topography as a continuous field $z_\zeta(\mathbf{x})$ with $\mathbf{x} = (x,y)^\top$ and write the inverse two-dimensional Fourier representation

$$z_\zeta(\boldsymbol{x}) = \frac{1}{(2\pi)^2} \int_{\mathbb{R}^2} \tilde{z}_\zeta(\boldsymbol{k}) \, e^{i\,\boldsymbol{k}\cdot\boldsymbol{x}} \, d^2\boldsymbol{k}, \tag{8}$$

with $\mathbf{k} = (k_x, k_y)^\top$ representing the wave vector and $\mathrm{k} = ||\boldsymbol{k}||$ as its absolute value. The band limitation imposed by $\zeta$ corresponds to a restriction $k_{\mathrm{min}} \leq k \leq k_{\mathrm{max}}(\zeta)$. The finite measurement field $L$ introduces, through windowing, an effective lower cutoff of $k_{\mathrm{min}} \sim 2\pi/L$.

For self-affine surfaces, the spectral roughness structure is characterized by the isotropic power spectral density (PSD) $C(k)$, which typically scales in the range $k_{\mathrm{min}} \ll k \ll k_{\mathrm{max}}$ as

$$C(k) \propto k^{-2(1+H)}, \tag{9}$$

with $H \in (0,1)$ denoting the Hurst exponent [28]. Through $H$, the fractal dimension of the function graph is linked to the embedding dimension: for surfaces represented as graphs $z(x,y) \subset \mathbb{R}^3$, one has $D = 3 - H$(with $2 \leq D \leq 3$), and for one-dimensional profiles $z(x) \subset \mathbb{R}^2$, correspondingly $D = 2 - H$ (with $1 \leq D \leq 2$). Thus, $H$ is directly coupled to the spectral roughness structure, while $\zeta$ (via $k_{\mathrm{max}}(\zeta)$ or $\lambda(\zeta)$) determines the scale bandwidth resolved in the topographic representation [28]

The scale-dependent $\gamma(\zeta)$ is not to be interpreted as a material constant of the boundary layer, but rather as an energetic weighting of the geometric degrees of freedom, whose magnitude is set by the elastic storage energy of the band-limited topography.

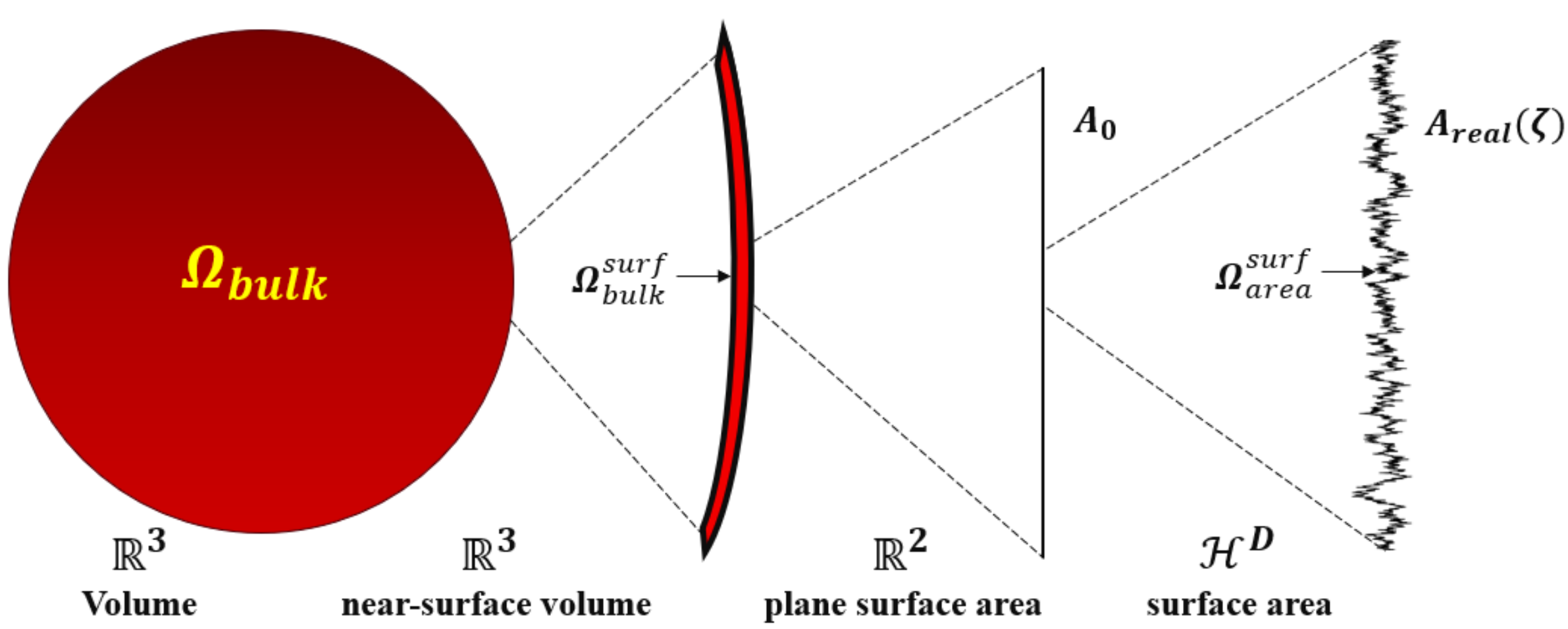

Fig. 3 Geometric decomposition of the surface subsystem into a near-surface volumetric part $\Omega_{\text{bulk}}^{\text{surf}} \subset \mathbb{R}^3$ and a scale-dependent areal part $\Omega_{\text{area}}^{\text{surf}}$, which transitions from the projected reference area $A_0 \subset \mathbb{R}^2$to the real surface $A_{\text{real}}(\zeta)$with an effective Hausdorff measure $H^D\,(2 < D < 3\,)$.

In Persson's full-contact reference case up to the cutoff $k_{\max}(\zeta)$, the maximal scale-dependent elastic storage energy per unit projected area is given by [15]

$$\frac{U_{el}^0(\zeta)}{A_0} = \frac{E^*}{4}\int_{k_0}^{k(\zeta)} k\, C(\boldsymbol{k})\; d^2\boldsymbol{k}\,. \tag{10}$$

Here, the half-space elastic response weighs each spectral mode by its wavenumber magnitude. Hence, the energy scale is directly tied to the directional gradient statistics encoded in the PSD.

In our approach, we retain Persson's separation of geometry (PSD) and elastic properties (via the effective modulus $E^*$), but reverse the assignment: the elastic energy stored in the half-space in Persson's reference picture is interpreted as the energetic scale of the topography subsystem, i.e., as a measure of the compression of the resolved geometric degrees of freedom under the elastic resistance of the contact. Thus, $\gamma(\zeta)$ is linked to $U_{\text{el}}^0(\zeta)$ through a reference calibration: in the limiting case of maximal activation, the reversible surface work of the areal subsystem is scaled such that it reproduces Persson's energy scale:

$$\Omega_{area}^{sub}(\zeta) \mathrel{\widehat{=}} U_{el}^0(\zeta), \tag{11}$$

fixes $\gamma^0(\zeta)$ as the reference value of the conjugate quantity.

Figure 3 illustrates the scale hierarchy from the volume $\mathbb{R}^3$ through the near-surface volumetric region to the projected reference area $A_0 \subset \mathbb{R}^2$ and to the real, scale-dependent surface $A_{\text{real}}(\zeta)$, whose contribution enters the potential formulation as its own term $\Omega_{\text{area}}^{\text{sub}}(\zeta)$. Accordingly, the potential formulation can be written as

$$\mathrm{d}\Omega_{\mathrm{surf}}^{\mathrm{sub}}(\zeta) = \mathrm{d}\Omega_{\mathrm{bulk}}^{\mathrm{sub}} + \mathrm{d}\Omega_{\mathrm{area}}^{\mathrm{sub}}(\zeta). \tag{12}$$

In the unloaded reference limit, which includes the removal of mechanical loading and sufficient relaxation of the relevant microscopic degrees of freedom, the coupled system relaxes, on the scales considered, into a state of minimal free energy. From the perspective of the surface description, this means that microscopic degrees of freedom of the boundary layer and geometric degrees of freedom of the topography jointly determine the scale-dependent partition function of the surface.

In statistical mechanics, the partition function $Z = \sum_i \exp(-\beta E_i)$ encodes the statistical weight of all accessible microstates at a given temperature; here, the microstates are the geometric configurations of the roughness at resolution $\zeta$, and the “temperature” is set by the elastic energy scale $\gamma(\zeta)$.

The associated scale-dependent surface area needs to be regarded as an ensemble quantity, whose representation will be specified in the following by a number $N(\zeta)$ of state representatives (ensemble quanta) that map the statistical asperity ensemble of the topography. We want to show that, for randomly rough surfaces, $\mathrm{d}\Omega_{\mathrm{area}}^{\mathrm{sub}}(\zeta)$ assumes the structure of a classical thermodynamic potential:

$$\mathrm{d}\Omega_{\mathrm{area}}^{\mathrm{sub}}(\zeta) = \gamma(\zeta)\mathrm{d}A_{\mathrm{real}} = \frac{1}{\beta_{\mathrm{topo}}(\zeta)}\,\mathrm{d}S_{\mathrm{topo}}(\zeta), \tag{13}$$

where $\beta_{\mathrm{topo}}(\zeta)$ emerges as the Lagrange multiplier of a maximum-entropy (MaxEnt) variation with constraints and is an effective inverse temperature in the sense of Öttingers and Grmela’s multiscale thermodynamics [25,29]. Specifically, $\beta_{\mathrm{topo}}(\zeta)$ is conjugated to the energetic weighting of the geometric degrees of freedom and arises when the elastic storage energy $U_{\mathrm{el}}^{0}(\zeta)$ (Eq. 10) is imposed as an energy constraint. It is not a bath temperature and does not require a kinetic interpretation, its meaning is fully determined by the chosen entropy functional $S_{\mathrm{topo}}(\zeta)$ and by the set of constraints defining the MaxEnt equilibrium on the reduced description level.

In the limiting case $\zeta \to 1$, no geometric degrees of freedom are resolved: $A_{\mathrm{real}} = A_0$, and the statistical formulation reduces to the classical shape equilibrium of Eq. (3)

$$U_{el}^{0}(\zeta) \mathrel{\widehat{=}} U_{el(topo)}^{0}(\zeta), \tag{14}$$

as discussed in [27], §159 for the equilibrium of a single crystal. The passage from this geometric limit to the full statistical formulation at finite $\zeta$ is the subject of the following chapter.

## 3. The topography ensembles.

We model the topography as a band-limited, stationary field $z_\zeta(\mathbf{x})$ whose bandwidth is set by the magnification $\zeta$ via $k_0 \le |\mathbf{k}| \le k(\zeta)$. Although the real surface area $A_{\mathrm{real}}(\zeta)$ may diverge as $\zeta \to \infty$, the elastically stored energy $U_{\mathrm{el}}^{0}(\zeta)$ converges to a finite limit for Hurst exponents $H > 1/2$ [15, 29]. In the reference case of full elastic contact between a rigid flat and an elastic half-space, the stored energy per unit nominal area reads

$$\frac{U_{\mathrm{el}}^{0}(\zeta)}{A_0} = \frac{E^*}{4} \int_{k_0 \le |\mathbf{k}| \le k(\zeta)} k \; C(\mathbf{k}) \; \mathrm{d}^2 k \tag{15}$$

where $k = |\mathbf{k}|$ is the magnitude of the wave vector and the elastic half-space response weights each spectral mode proportionally to $k$. The second expression assumes isotropy, $C(\mathbf{k}) = C(k)$.

This purely spectral description provides a complete statistical characterization if and only if the field is stationary and Gaussian [13]. For a Gaussian process, the autocovariance, and hence the PSD, determines the entire statistical structure. In particular, the covariance between height and gradient vanishes,

$$\mathrm{Cov}(z,\, \partial_x z) = \left. \frac{\partial R(\tau)}{\partial \tau} \right|_{\tau=0} = 0, \tag{16}$$

and for a jointly Gaussian field this zero covariance implies statistical independence of height and gradient. The spectral moments $m_0$, $m_2$, $m_4$ of $C(k)$, with $m_n = \int k^n C(k) d^2 k$, then provide a complete description [13].

Real surfaces, however, frequently violate Gaussian behavior. Stout et al. [31] showed in a running-in experiment that the height distribution $\rho(z)$ develops a pronounced negative skewness ($R_{\mathrm{sk}} < 0$), so the field is no longer Gaussian. Pérez-Ràfols and Almqvist [32] confirmed this consequence numerically: for self-affine surfaces with identical PSD but Weibull-distributed heights, the contact stiffness deviates systematically from the linear relation established for Gaussian surfaces. The PSD as a second-order descriptor is therefore no longer sufficient to predict the contact mechanics.

A second practical limitation concerns anisotropy. Persson's theory assumes isotropic topographies, yet real technical surfaces are typically anisotropic. Following the method described by Pérez-Ràfols and Ciavarella [33], it is possible to generate anisotropic roughness fields and compare two topographies with comparable $S_q \approx 0.3$ $\mu$m and Hurst exponent $H \approx 0.8$, as shown in Figure 4. The area-averaged root-mean-square gradient $S_{\mathrm{dq}}$ (defined in ISO 25178 [35]) of the anisotropic surface is reduced compared to the isotropic case, because the spectral modes are concentrated along a preferred direction. The spatial gradient field $P(\partial z/\partial x, \partial z/\partial y)$ in Figure 4 makes this visible: for the anisotropic surface, the gradients cluster along a single axis. Projecting the gradient field onto a direction $\varphi$ and choosing $\varphi^*$ such that the directional gradient is maximal, both surfaces possess the same directional gradient, which we denote $R_{\mathrm{dq}}$ in accordance with ISO 21920 [34].

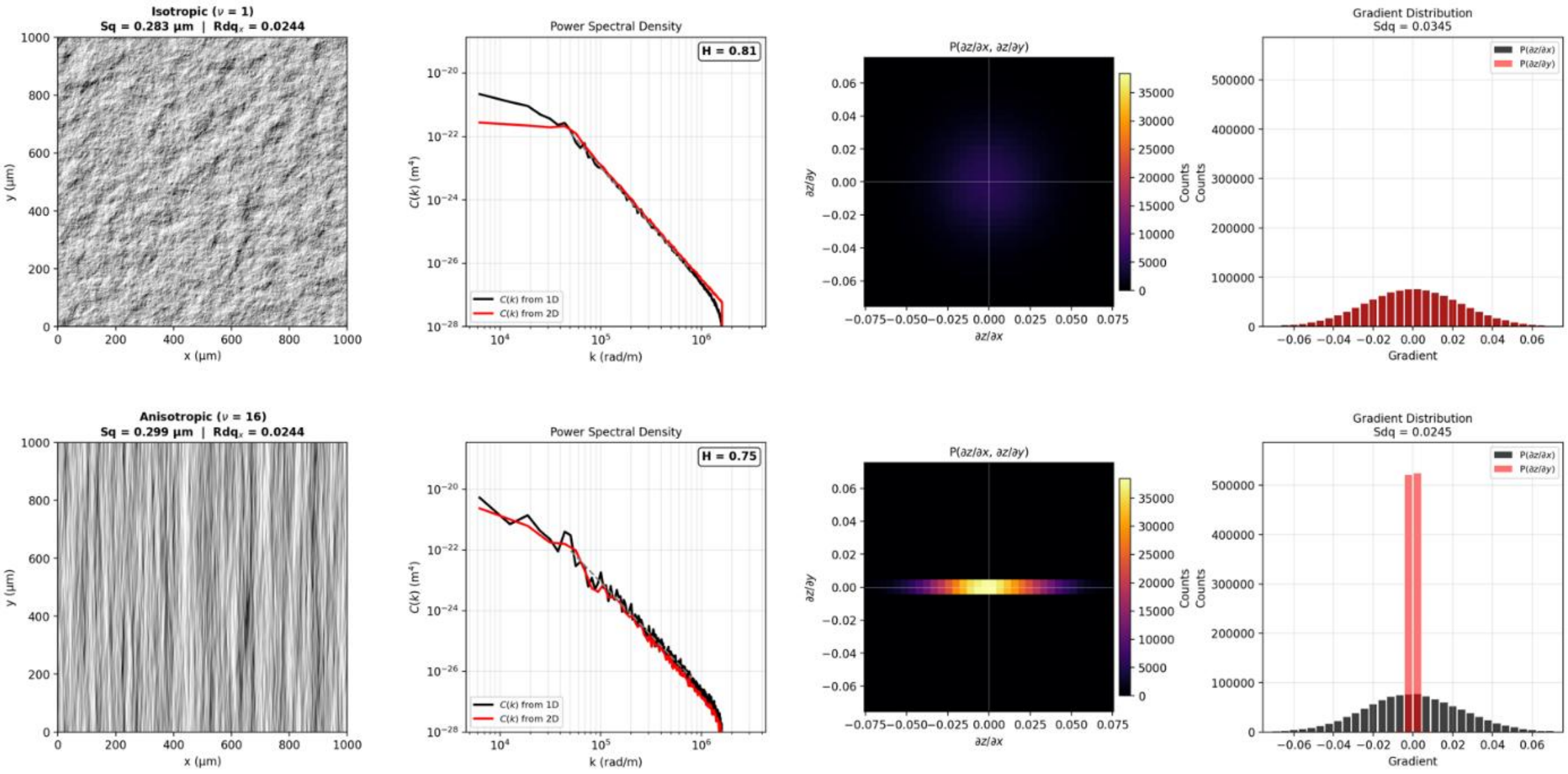

**Figure 4.** Isotropic (β = 1, top) and anisotropic (β = 16, bottom) topography with $S_q \approx 0.3\ \mu$m, $H \approx 0.8$, generated following [33]. From left: height map, PSD, gradient field $P(\partial z / \partial x, \partial z / \partial y)$, marginal gradient distributions. Despite different $S_{dq}$ (0.0345 vs. 0.0245), both surfaces share $R_{dq} = 0.0244$ along the direction of maximum gradient.

This observation motivates a reduction in dimension: rather than considering the full two-dimensional field $z(x, y)$, we extract profiles $z(x)$ along the direction of maximum gradient. The directional gradient $R_{\mathrm{dqx}}$ captures the gradient information most relevant to elastic contact, which is diluted in the area-averaged $S_{\mathrm{dq}}$ by spatial averaging. For isotropic surfaces, the two directional contributions are identical, so

$$S_{\mathrm{dqx}}^2 = \langle(\partial_x z)^2\rangle + \langle\left(\partial_y z\right)^2\rangle = 2\langle(\partial_x z)^2\rangle = 2\,R_{\mathrm{dqx}}^2. \tag{17}$$

We therefore project the elastic surface energy onto profiles along $\varphi^*$. The projection replaces the isotropic PSD $C(k)$ by the projected profile PSD $C_{\varphi^*}(\mathbf{k})$ along the direction of maximum gradient. The integral itself retains its two-dimensional form over $\mathrm{d}^2k$, because the projection does not eliminate the transverse wave vector $k_y$ but merely stretches the spectral distribution along $\varphi^*$. The measure $\mathrm{d}^2k$ is preserved, and with it the dimension of the integral. The projected full-contact reference energy then reads

$$\frac{U_{\mathrm{el}}^0(\zeta)}{A_0} = \frac{E^*}{4}\int k\ C_{\varphi^*}(\mathbf{k})\ \mathrm{d}^2k. \tag{18}$$

Having reduced the gradient statistics to profiles $z(x)$ along the direction of maximum gradient while retaining the two-dimensional energy integral of Eq. (18), we now return the roughness field introduced by Nayak [13] to its origin: Longuet-Higgins' theory of randomly propagating ocean waves [36, 37]. The band limitation $k_0 \leq k \leq k(\zeta)$ renders the information density of the field finite. Nayak transferred this framework to technical solid surfaces by interpreting the surface profile as a spatial realisation of a band-limited random field, thereby mapping the temporal wave statistics onto the spatial statistics of the topography.

For a stationary Gaussian roughness field, Persson's theory represents the quadratic reference limit of such a band-limited random process. The scale parameter $\zeta$ restricts the spectral bandwidth to $k \leq \zeta\, k_0$, and by Gabor's theorem [37] the resulting signal is discretised into $N = 2L/\lambda_s$ independent degrees of freedom, called logons. This discretisation implies a signal-theoretic uncertainty

$$\Delta r\ \Delta k \gtrsim 1/2 \tag{19}$$

Parseval's theorem guarantees that the total energy is invariant under Fourier transformation: the energy computed from the spatial signal $z(r)$ equals the energy computed from the power spectral density $C(k)$. In the Gaussian limit, all statistical moments can be extracted from the spectrum alone, and the marginal distributions of height and gradient are independent. The spectral moments $m_0$, $m_2$ and $m_4$ then suffice to characterize the process completely, and Persson's diffusion equation for the contact stress distribution $P(\sigma, \zeta)$ [15,30],

$$\frac{\partial P(\sigma,\zeta)}{\partial \zeta} = D(\zeta)\,\frac{\partial^2 P(\sigma,\zeta)}{\partial \sigma^2}, \quad (20)$$

propagates this spectral information through the coarse-graining parameter $\zeta$, where $D(\zeta)$ is determined entirely by the spectral content and elastic properties at magnification $\zeta$ [15, 30] This equivalence, however, holds exclusively for Gaussian processes. For non-Gaussian surfaces, the marginals are no longer independent, and the power spectrum alone cannot resolve the correlations between height and gradient that govern elastic energy.

To make these correlations accessible, we return to the spatial domain and identify the local degrees of freedom. Along a band-limited profile, each position $r$ carries exactly one vertical displacement, which serves as the generalised coordinate $\mathfrak{q}$. The conjugate quantity arises naturally from the spatial structure: as the magnification $\zeta$ increases, additional spectral modes are resolved and each local height $\mathfrak{q}(r)$ is refined relative to its neighbours. The local gradient $d\mathfrak{q}/dr$ captures this refinement. It measures how rapidly the surface changes along the spatial coordinate at a given resolution. Crucially, this is a structural rate of change, not a temporal or kinematic one. It quantifies the geometrical complexity that each resolution step reveals, not a velocity in the mechanical sense.

We decompose the elastic reference energy (Eq. 18) into two quadratic contributions, one depending on the local gradient, the other on the local height and write the scale-dependent structural generator

$$G = K - B \quad (21)$$

in formal structural mathematical analogy to a Lagrangian. The gradient term

$$K_{\mathrm{topo}} = \Lambda^{-1}\mathrm{E}^*\,\mathfrak{p}^2, \qquad (K_{\mathrm{topo}}) = \left(\frac{\mathrm{J}}{\mathrm{m}^2}\right) \quad (22)$$

represents the elastic energy stored in the half-space due to the local surface slope $\mathfrak{p} = d\mathfrak{q}/dr$. The prefactor $\Lambda$ couples the dimensionless gradient to the elastic stiffness $E^*/A_0$ of the average contact mode $\Lambda = \int k\; C_{\varphi^*}(\mathbf{k})\; \mathrm{d}^2k / \int C_{\varphi^*}(\mathbf{k})\; \mathrm{d}^2k\;\; (m^{-1})$. It measures the rate at which the half-space storage energy increases with the square of the surface gradient, i.e. with the local surface enlargement. In the ensemble average, $\langle K_{\mathrm{topo}} \rangle = \left( U_{\mathrm{el}}^0(\zeta)/A_0 \right) m_2$, which recovers the energy scale of Persson's spectral integral (Eq. 18). The height term

$$B_{\mathrm{topo}}(\mathfrak{q}) = \Lambda E^*\,(q_{\mathrm{max}} - \mathfrak{q})^2, \qquad (B_{\mathrm{topo}}) = \left(\frac{\mathrm{J}}{\mathrm{m}^2}\right) \quad (23)$$

represents the elastic energy associated with the pressure-induced indentation of a logon at height $\mathfrak{q}$ relative to the contact level $q_{\mathrm{max}}$. The quantity $(q_{\mathrm{max}} - \mathfrak{q})$ is the indentation depth; its square reflects the quadratic scaling of elastic half-space energy with penetration. At the contact level ($\mathfrak{q} = q_{\mathrm{max}}$), $B = 0$ and the entire energy resides in the gradient channel $K$. Deep below the contact level, $B$ dominates: large external work is required to bring the material into contact. The origin of the height coordinate may be placed on the Abbott–Firestone bearing curve without loss of generality. Both $K$ and $B$ carry the same prefactor $E^*/A_0$ because both represent elastic deformation of the same half-space, $K$ through surface slope (flank geometry) and $B$ through indentation (pressure work). Their sum is calibrated by

$$\langle K_{\text{topo}} + B_{\text{topo}} \rangle = \frac{U_{\text{el}}^{0}(\zeta)}{A_0}, \qquad (24)$$

so that the mean structural energy, averaged over all logons, reproduces Persson's full-contact reference energy per unit projected area.

The Legendre transformation and the canonical structure it induces, conjugate variables, a symplectic bracket, a partition into kinetic and potential contributions, are mathematical properties of any convex decomposition of an energy functional [38]. Classical dynamics is one realisation of this structure, thermodynamics another. Here we use a third: $G$ defines, via Legendre transformation, the conjugate gradient

$$\mathfrak{p} = \frac{d\mathfrak{q}}{dr}, \qquad (25)$$

which defines the conjugate pair $(\mathfrak{q}, \mathfrak{p})$ and guarantees the geometrical canonical bracket $\{\mathfrak{q}, \mathfrak{p}\} = 1$. The resulting structural energy functional

$$E(\mathfrak{q}, \mathfrak{p}) = K_{\text{topo}}(\mathfrak{p}) + B_{\text{topo}}(\mathfrak{q}) = E^{*} \cdot [\Lambda^{-1}\mathfrak{p}^2 + \Lambda(\mathfrak{q}_{\text{max}} - \mathfrak{q})^2] \qquad (26)$$

has the mathematical form of an isotropic harmonic oscillator in the topographic phase space. No equation of motion is associated with $G$, the oscillator structure is purely energetic. For a single sinusoidal mode $z(r) = A\sin(kr)$, the identity $\mathfrak{q}^2 + (\mathfrak{p}/k)^2 = A^2$ maps each profile point onto an ellipse in $(\mathfrak{p}, \mathfrak{q})$, confirming the quadratic energy partition analytically (cf. Fig. 5).

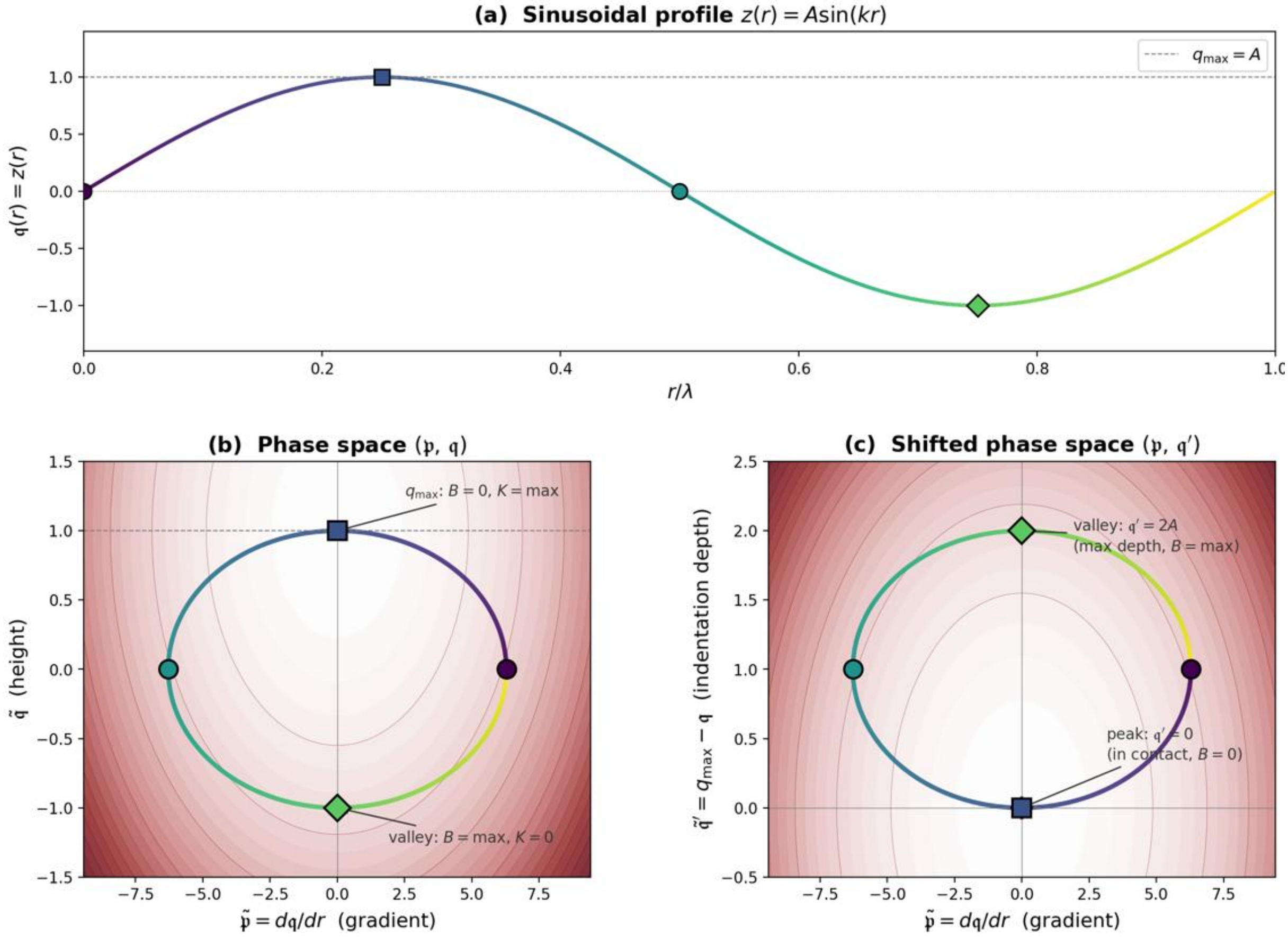

**Figure 5.** Topographic phase space for a sinusoidal profile $z(r) = A\sin(kr)$. (a) Profile with characteristic positions. (b) Normalised phase space $(\tilde{\mathfrak{p}}, \tilde{\mathfrak{q}})$: the trajectory traces an ellipse through energy contours $E = E^{*}[\Lambda^{-1}\mathfrak{p}^2 + \Lambda(q_{max} - \mathfrak{q})^2]$. At the peak, B=0; at the valley, K=0. (c) Shifted coordinates $\mathfrak{q}' = q_{max} - \mathfrak{q}$ (indentation depth): $\mathfrak{q}' = 0$ at the contact level, $\mathfrak{q}' = 2A$ at maximum indentation.

In the Gaussian limit, $E(\mathfrak{q}, \mathfrak{p})$ is quadratic in both variables. The joint density factorises as $\rho_G(\mathfrak{q}, \mathfrak{p}) = \mathcal{N}(0, \sigma_{\mathfrak{q}}) \cdot \mathcal{N}(0, \sigma_{\mathfrak{p}})$, and the variances reduce to $\sigma_{\mathfrak{q}}^2 = m_0$ and $\sigma_{\mathfrak{p}}^2 = m_2$. Both underlying moments are evaluated from the profile along $\varphi^{*}$. Since Persson's spectral description thus represents the Gaussian limit of the topographic state, we define the general state through the canonical partition function

$$Z = \sum_{i} \exp\left(-\beta\, E_i\right), \qquad (27)$$

where $E_i = E(\mathfrak{q}_i, \mathfrak{p}_i)$ is the energy of the $i$-th phase-space cell and $\beta$ is the Lagrange multiplier conjugate to the mean energy $\langle E \rangle = U_{\mathrm{el}}/A_{\mathrm{real}}$, the actual elastic energy under partial contact, as distinct from the full-contact reference $U^{0}_{\mathrm{el(topo)}}/(m_2\, A_0)$ projected on the real surface area of the topography.

The energy cells of this phase space require a state-independent discretization that does not presuppose the form of the distribution to be analyzed. We therefore orient the binning at the uncorrelated Gaussian process as a reference case. For normally distributed data, Scott's rule [39] minimizes the integrated mean squared error of the histogram estimator and yields a bin width

$$c_w = \frac{3.49\,\sigma}{N^{1/3}}, \qquad (28)$$

which we apply independently to both coordinates: $\Delta q = 3.49\,\sigma_q/N^{1/3}$ for the height axis and $\Delta p = 3.49\,\sigma_p/N^{1/3}$ for the gradient axis. The factor 3.49 is exact for the Gaussian reference. The grid is centred symmetrically on zero with an odd number of bins in each direction, so that the central tile straddles the origin. For non-Gaussian surfaces the binning remains unchanged, ensuring that deviations from Gaussianity become visible in the occupation pattern $\rho(q_i, p_i)$ rather than being absorbed into discretization.

The extent of the event space follows Seewig [40], who shows that for sample sizes of order $N \sim 10^4$ the statistically stable range of the height distribution is bounded by $\pm 5\sigma$. Using discrete bin occupancies $\{n_i\}$ with $\rho_i := n_i/N$ and $\sum_i \rho_i = 1$, the standard combinatorial argument via the Stirling approximation yields the extensive topographic entropy

$$S_{\text{topo}} = \ln W = \ln \frac{N!}{\prod_i n_i\,!} \approx -N \sum_i \rho_i \ln \rho_i. \qquad (29)$$

From this point, the equilibrium distribution follows from a maximum-entropy variation in the sense of Jaynes [41]. We maximize $S_{\text{topo}}$ subject to normalisation and the prescribed energy constraint, which leads to the canonical equilibrium density

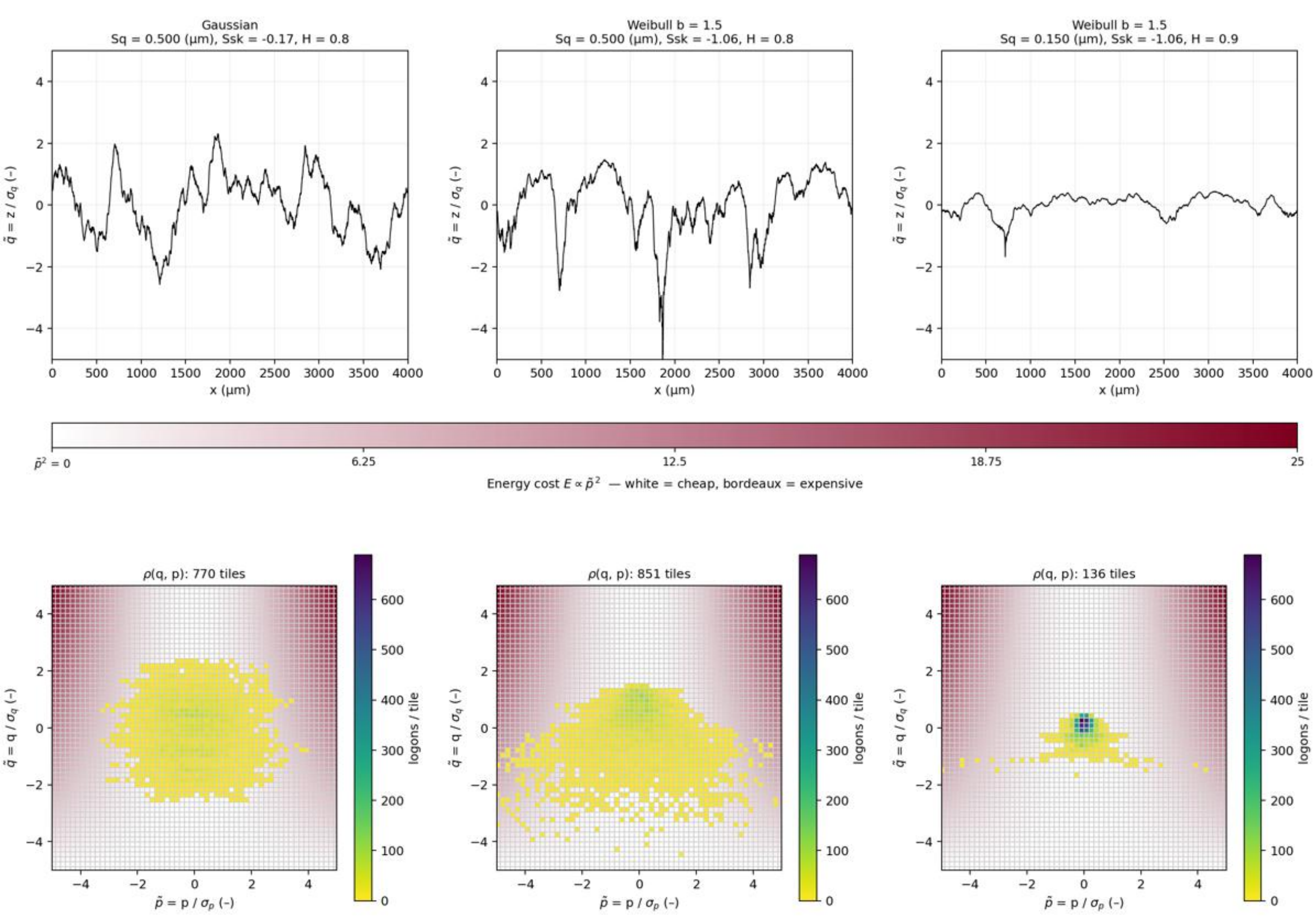

**Figure 6.** Phase-space representation of three self-affine surface states discretised on the common Scott grid (Eq. 28). Upper row: normalised profiles $\tilde{q} = z/\sigma_q$. Lower row: phase-space density $\rho(q, p)$ with energy cost $E \propto \tilde{p}^2$ encoded as bordeaux background (white = low cost, dark = high cost). Occupied cells coloured by logon count (viridis). Calibrated at E* = 115 GPa (100Cr6).

$$\rho^*(\mathfrak{q}_i, \mathfrak{p}_i) = \frac{1}{Z} \exp\big(-\beta\, E(\mathfrak{q}_i, \mathfrak{p}_i)\big), \qquad (30)$$

recovering the Boltzmann form with $\beta = \beta_{\text{topo}}$ as the topographic inverse temperature [29]. In the Gaussian limit, Eq. (30) reduces to the product of two independent Gaussians with variances $m_0$ and $m_2$, recovering Persson's spectral description exactly. For non-Gaussian surfaces, $\rho^*(\mathfrak{q}_i, \mathfrak{p}_i)$ is no longer factorisable, and the correlation between gradient class and height class determines the actual elastic energy stored under contact loading.

We have thus shown that Persson's spectral approach for the elastic energy stored in the half-space can be generalized by maximizing the entropy of the statistically distributed topographic degrees of freedom at a given scale $\zeta$. This yields a canonical Boltzmann distribution over the topographic phase space whose Gaussian limit reproduces Persson's spectral description. The Lagrange Multiplier associated with the energy constraint,

$$\beta_{\text{topo}} = \frac{\partial S_{\text{topo}}}{\partial \langle E \rangle}, \qquad (31)$$

can, within Grmela's multiscale thermodynamics [29], be identified as the multiplicative inverse of a mesoscopic temperature $T_{\text{topo}}$. This quantity characterises the sensitivity of the topographic entropy to changes in the stored elastic energy at fixed scale and thereby incorporates the state function introduced in Chapter 2,

$$d\Omega_{\text{area}} = \frac{1}{\beta_{\text{topo}}}\, dS_{\text{topo}}. \qquad (32)$$

To illustrate the framework, Figure 6 compares three self-affine surface states whose phase-space densities $\rho(q, p)$ are discretised on the common Scott grid (Eq. 28,29). The energy cost of each cell follows from Eq. (26): the kinetic contribution $K_{\text{topo}} \propto p^2$ increases quadratically with the canonical gradient $p$ (Eq. 22), assigning high cost to cells with steep local slopes, while the potential contribution $B_{\text{topo}}(q)$ weights cells by their position in the bearing area, since only the upper portion of the height distribution participates in contact loading. The product of both contributions is encoded as a bordeaux background shading from white (negligible cost) to dark (maximum stored energy per cell). Occupied cells are coloured by their logon count.

**Table 1.** State properties of the three surface types at $E^*$ = 115 GPa. The intensive entropy $s = S_{\text{topo}}/N$ is the topographic entropy per logon.

| | Unit | Gaussian | Plateau | Smooth | Eq. |
|---|---|---|---|---|---|
| $S_q$ | (µm) | 0.500 | 0.500 | 0.150 | — |
| $S_{sk}$ | (–) | −0.17 | −1.06 | −1.06 | — |
| $H$ | (–) | 0.8 | 0.8 | 0.9 | — |
| $R_{dq}$ | (–) | 0.0152 | 0.0261 | 0.0054 | — |
| Occ. tiles | (–) | 787 | 875 | 144 | (28,29) |
| $s = S_{\text{topo}}/N$ | (–) | 6.67 | 6.43 | 3.82 | (29) |
| $\langle E \rangle$ | (J/m²) | 200 | 136 | 14 | (24) |

The pairwise finite differences $\Delta s/\Delta\langle E\rangle$ yield the mean inverse temperature $\bar{\beta} = \Delta s/\Delta\langle E\rangle$ along each path in state space (Eq. 31). Because $s$ is the entropy per logon, the resulting topographic temperature $T_{topo} = 1/\bar{\beta}$ carries the dimension J/m², the same as the elastic energy density $U_{el}^{0}/A_{0}$. This is dimensionally required by $d\Omega_{area} = T_{topo}\, ds$. Table 2 lists the resulting temperatures.

**Table 2.** Pairwise state differences and topographic temperature.

| **State pair** | **Δs** | $\Delta\langle E\rangle$ (J/m²) | $\bar{\beta}$ (m²/J) | $T_{topo}$ (J/m²) |
|---|---|---|---|---|
| Gaussian ↔ Plateau | 0.24 | 64 | 0.004 | 266 |
| Plateau ↔ Smooth | 2.61 | 122 | 0.021 | 47 |
| Gaussian ↔ Smooth | 2.85 | 186 | 0.015 | 65 |

The Gaussian surface is topographically hot ($T_{topo} \approx 266$ J/m² between Gaussian and Plateau): each unit of stored elastic energy supports many geometric microstates. The Smooth surface is cold ($T_{topo} \approx 47$ J/m² between Plateau and Smooth): its compact phase-space distribution binds entropy tightly to energy. The overall path Gaussian ↔ Smooth yields $T_{topo} = 65$ J/m², dominated by the cold Plateau ↔ Smooth segment which accounts for 92 % of $\Delta s$. The temperature decreases monotonically with decreasing $\langle E\rangle$, confirming the concavity of the $s(\langle E\rangle)$ curve and thus thermodynamic stability. In the limit $\zeta \rightarrow 1$, no geometric degrees of freedom are resolved: $s = 0$, $\langle E\rangle = 0$, and the $s(\langle E\rangle)$ curve passes through the origin. This limit corresponds to the defect-free single crystal whose shape equilibrium was identified in Chapter 2 (Eq. 14) as the classical reference following Landau and Lifshitz [27, §159]. The single crystal thus represents the absolute zero of the topographic temperature scale, from which the Smooth, Plateau and Gaussian states ascend in the order of increasing geometric complexity.

## 4. Experimental validation

The preceding chapters established a phase-space description $\rho(q,p)$ of band-limited topographies whose Gaussian limit recovers Persson's spectral theory (Chapter 3, Fig. 6). We now confront this framework with measured surfaces from a tribological running-in experiment and show that the topographic temperature $T_{topo}$ provides a consistent scale on which different running-in mechanisms become quantitatively comparable.

Two AISI 52100 (100Cr6) steel discs, one directionally ground (N1V2) and one honed (NH1), were subjected to identical running-in protocols in a Mini Traction Machine [43] (ball-on-disc, 23 N, entrainment speeds 10 to 2 000 mm/s, bath temperatures 40 to 100 °C, PAO-based lubricants with corrosion inhibitor, sealed test pot). The ground surface reached steady-state friction after approximately 55 h, the honed surface after approximately 11 h. Surface topographies were measured before and after running-in using a Nanofocus µsurf confocal microscope (multi-pinhole technique [42], 50× objective, NA = 0.8, field 320 × 320 µm², lateral sampling ≈ 0.23 µm). Raw topographies were processed by second-order polynomial form removal followed by an isotropic S-filter at $\lambda_s$ = 2.5 µm. The directional gradient $p$ was computed with a seven-point stencil along the direction of maximum gradient. The phase-space discretisation uses $N$ = 65 536 logons with Scott bin widths (Eq. 28) referenced to the ground initial surface V0626 (128 × 128 tiles, ±5$\sigma$ range, energy calibration $U_{el}^0/A_0$ = 2 819 J/m² at $E^*$ = 115.4 GPa). Because all four surfaces are discretised on this common grid, the intensive entropy $s = S_{topo}/N$ and the mean energy $\langle E \rangle$ of every state are evaluated on a single scale. Table 3 summarises the surface parameters after S-filtering.

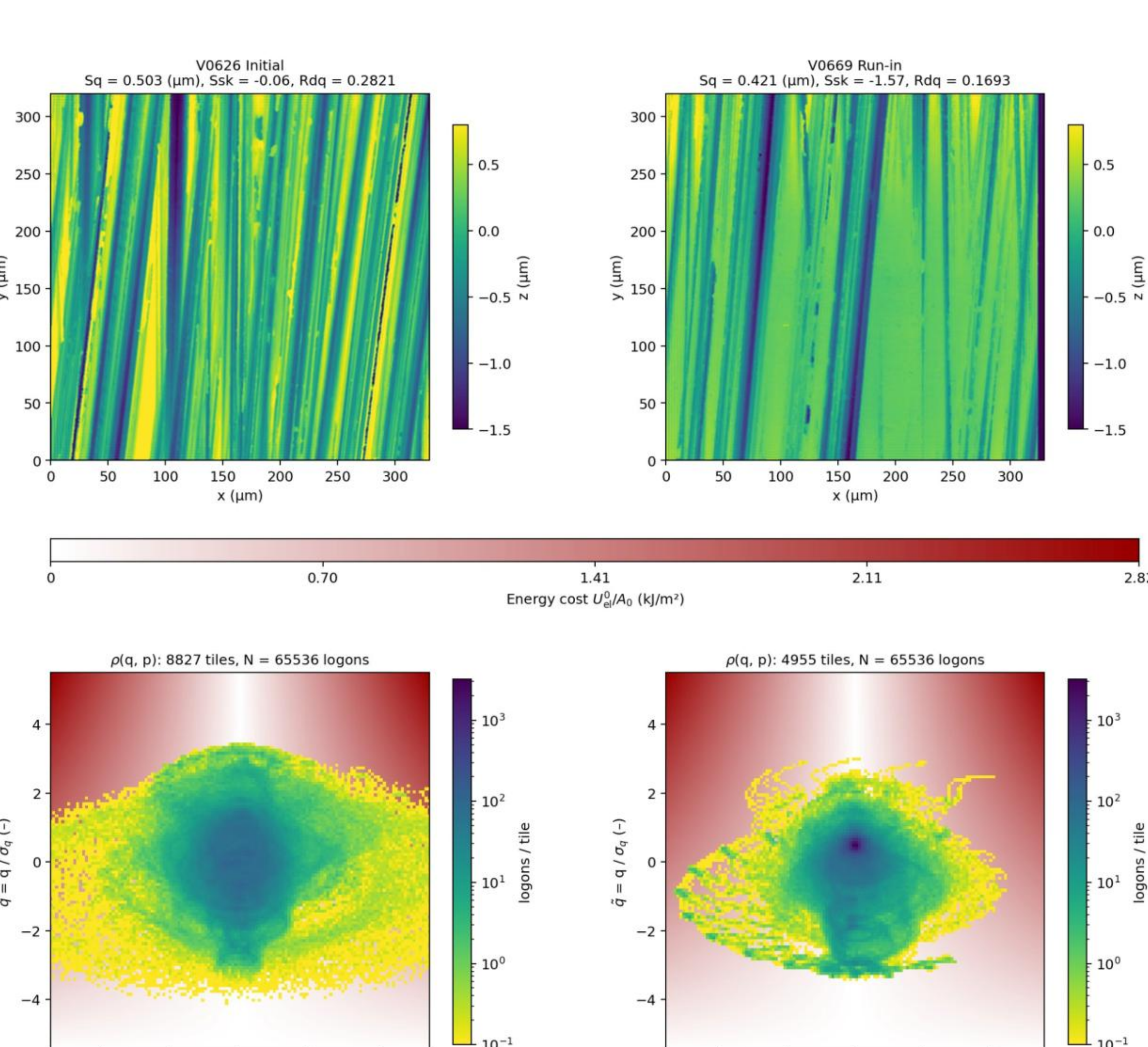

**Figure 7.** Phase-space analysis of the ground disc (100Cr6), same reference grid as Fig. 6. Left: V0626 initial (8 827 tiles, Sq = 0.503 µm, Ssk = −0.06, Rdqx = 0.282). Right: V0669 run-in (4 955 tiles, Sq = 0.421 µm, Ssk = −1.57, Rdqx = 0.169). Upper row: topography maps (320 × 320 µm²). Lower row: phase-space density ρ(q, p) with energy cost encoded as bordeaux background. Running-in contracts the occupied domain as logons migrate from high-cost flanks to the low-cost centre.

**Table 3.** Surface texture parameters after S-filtering ($\lambda_s$ = 2.5 μm). Uncertainties: standard error from line-by-line evaluation

| **Surface** | $S_q$ (μm) | $S_{sk}$ (–) | $S_{dq}$ (–) | $R_{dqx}$ (–) | u($S_q$) | u($S_{sk}$) |
|---|---|---|---|---|---|---|
| V0626 Ground init. | 0.503 | −0.06 | 0.292 | 0.282 | 0.001 | 0.01 |
| V0669 Ground run-in | 0.421 | −1.57 | 0.177 | 0.169 | 0.001 | 0.01 |
| V0086 Honed init. | 0.131 | −1.70 | 0.064 | 0.024 | 0.001 | 0.02 |
| V0555 Honed run-in | 0.198 | −2.79 | 0.085 | 0.042 | 0.002 | 0.02 |

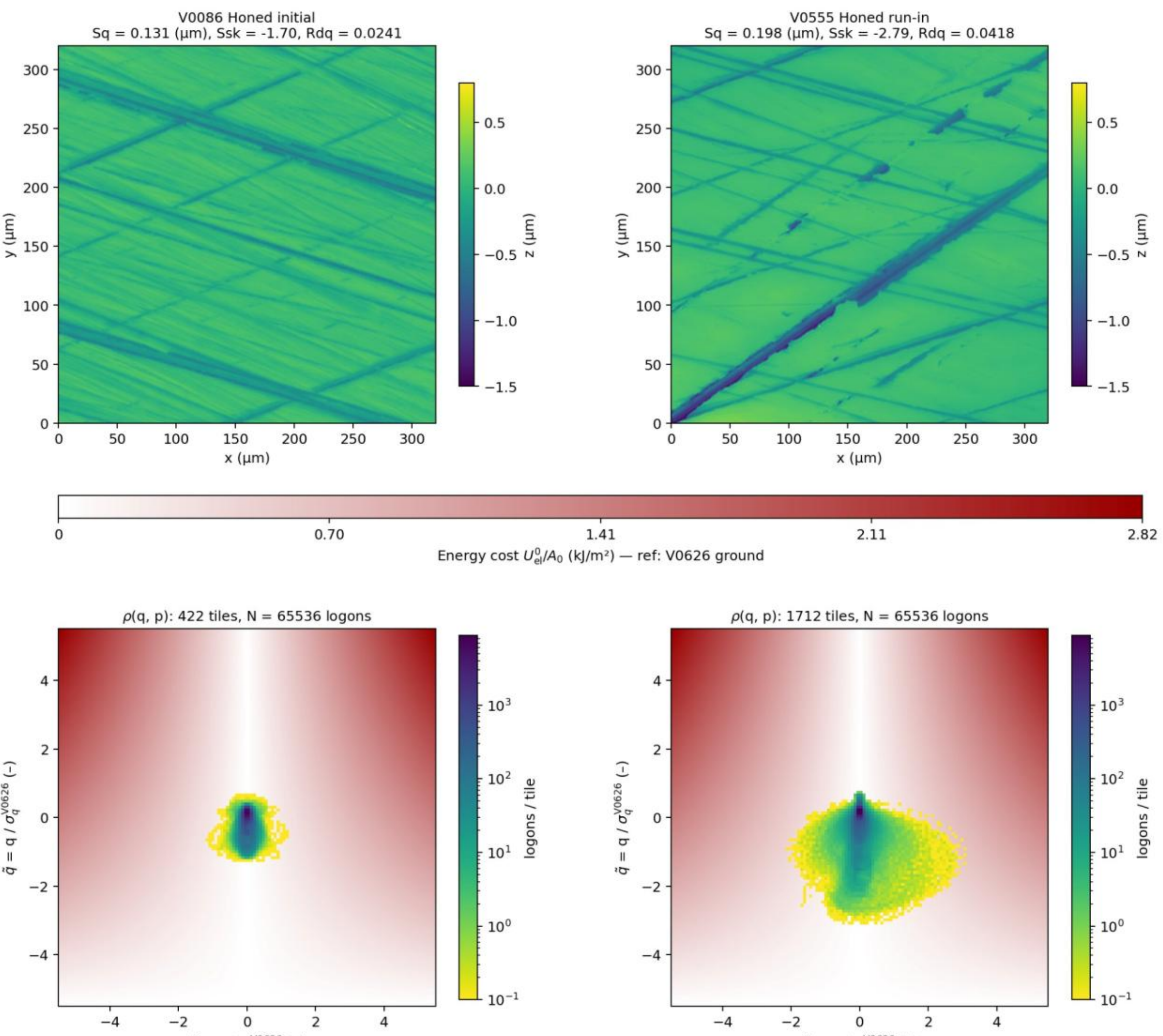

**Figure 8.** Phase-space analysis of the honed disc (100Cr6), same reference grid as Fig. 7. Left: V0086 initial (422 tiles, Sq = 0.131 μm, Ssk = −1.70, Rdqx = 0.024). Right: V0555 run-in (1 712 tiles, Sq = 0.198 μm, Ssk = −2.79, Rdqx = 0.042). Upper row: topography maps (320 × 320 μm²). Lower row: phase-space density ρ(q, p) with energy cost encoded as bordeaux background. Running-in expands the occupied domain as tribological loading relatively deepens grooves and populates previously unoccupied phase-space cells.

The ground initial surface V0626 is nearly Gaussian ($S_{sk}$ = −0.06) with the largest directional gradient ($R_{dqx}$ = 0.282) and spreads over 8 827 phase-space tiles. After running-in (V0669), $S_q$ decreases moderately (0.503 → 0.421 µm), but $R_{dqx}$ drops by 40 % and $S_{sk}$ shifts to −1.57. The occupied tiles contract to 4 955: logons migrate from the high-cost flanks to the low-cost centre, forming a plateau-like bearing surface (Fig. 8). The honed initial surface V0086 is already compact (422 tiles, $S_{sk}$ = −1.70, $R_{dqx}$ = 0.024). After running-in (V0555), $S_q$ rises to 0.198 µm, $R_{dqx}$ nearly doubles, $S_{sk}$ deepens to −2.79, and the occupied domain expands to 1 712 tiles. Tribological loading on the honed surface expands the volume of required state-tiles.

Table 4 lists the thermodynamic state variables computed from the phase-space densities (Eqs. 29 to 32). Table 5 lists the topographic temperatures for all six pairwise comparisons. The secant slope $\bar{\beta} = \Delta s/\Delta\langle E\rangle$ between any two states yields the mean inverse temperature on that segment of the $s(\langle E\rangle)$ curve; $T_{topo} = 1/\bar{\beta}$ assigns each pair a position on a common temperature scale anchored to the V0626 reference grid.

**Table 4.** Topographic state variables (Eqs. 29 to 32). Reference: V0626.

| **Surface** | **Tiles** | $s = S_{topo}/N$ | $\langle E\rangle$ (J/m²) |
|---|---|---|---|
| V0626 Ground init. | 8 827 | 7.53 | 170.1 |
| V0669 Ground run-in | 4 955 | 6.24 | 91.6 |
| V0086 Honed init. | 422 | 3.51 | 17.5 |
| V0555 Honed run-in | 1 712 | 3.63 | 20.2 |

The temperature ordering is monotonic and consistent with the synthetic examples of Chapter 3. The ground running-in path (V0626 → V0669) traverses the hot region of the $s(\langle E\rangle)$ curve at $T_{topo}$ = 61 J/m², comparable to the synthetic Gaussian ↔ Smooth temperature of 65 J/m²: the nearly Gaussian roughness of V0626 distributes elastic energy broadly across many phase-space cells, and running-in removes steep asperity flanks, releasing both energy and entropy. The honed running-in path (V0086 → V0555) occupies the cold region at $T_{topo}$ = 24 J/m²: the compact bearing area concentrates logons in low-energy cells, and running-in adds a small amount of geometric complexity as grooves deepen. The cross-comparisons between ground and honed surfaces ($T_{topo}$ = 27 to 38 J/m²) fill the gap between the two running-in paths and confirm the concavity of the $s(\langle E\rangle)$ curve, i.e. thermodynamic stability. Both running-in paths yield $T_{topo} > 0$, confirming the correct sign of the entropy/energy coupling.

**Table 5.** Pairwise state differences and topographic temperature $T_{topo} = 1/\bar{\beta}$.

| **State pair** | **Δs** | $\Delta\langle E \rangle$ (J/m²) | $\bar{\beta}$ (m²/J) | $T_{topo}$ (J/m²) |
|---|---|---|---|---|
| V0626 ↔ V0669 (Ground run-in) | 1.29 | 78.5 | 0.016 | 61 |
| V0626 ↔ V0086 (Ground ↔ Honed) | 4.01 | 152.6 | 0.026 | 38 |
| V0626 ↔ V0555 (Ground ↔ Hon. r-i) | 3.90 | 149.9 | 0.026 | 38 |
| V0669 ↔ V0086 (Gr. r-i ↔ Honed) | 2.73 | 74.1 | 0.037 | 27 |
| V0669 ↔ V0555 (Gr. r-i ↔ Hon. r-i) | 2.61 | 71.4 | 0.037 | 27 |
| V0086 ↔ V0555 (Honed run-in) | 0.11 | 2.7 | 0.043 | 24 |

## 5. Conclusion and outlook

This work introduced a thermodynamic state description of rough surfaces based on a canonical phase space $(\mathfrak{q}, \mathfrak{p})$ whose coordinates couple height and local gradient at a given scale. The intensive entropy $s = S_{\text{topo}}/N$ removes the dependence on the number of logons and yields a topographic temperature $T_{\text{topo}} = \Delta\langle E \rangle / \Delta s$ with the dimension J/m², the same as the elastic energy density. The $s(\langle E \rangle)$ curve is concave and passes through the origin, which corresponds to the defect-free single crystal as the absolute zero of the topographic temperature scale. The framework was validated experimentally by tracking the running-in of two differently finished 100Cr6 steel surfaces. The ground surface contracts in phase space ($\Delta s < 0$, $\Delta\langle E \rangle < 0$) at $T_{\text{topo}} = 61$ J/m², the honed surface expands ($\Delta s > 0$, $\Delta\langle E \rangle > 0$) at $T_{\text{topo}} = 24$ J/m². Both paths, together with the cross-comparisons between ground and honed states, define a monotonic temperature ordering that is consistent with the synthetic examples ($T_{\text{topo}} = 65$ J/m² for the Gaussian ↔ Smooth path). The formalism thus captures opposing running-in mechanisms through a single thermodynamic scale.

A critical limitation concerns the metrological uncertainty of the gradient coordinate $\mathfrak{p}$. Since the elastic energy functional (Eq. 26) weights the gradient quadratically, the energy scale is governed by $R_{\text{dqx}}$, which enters both the Scott binning (Eq. 28) and the energy calibration (Eq. 17). However, the mean square gradient is proportional to the second spectral moment $m_2$ of the PSD, which accumulates its contribution predominantly from the high-frequency end of the spectrum. As Persson [15] shows, the convergence of $m_2$ for self-affine surfaces depends on the Hurst exponent $H$: for $H < 1$, the integral over $k^2C(k)$ diverges logarithmically as the upper cutoff $k_{\max}$ increases, making the gradient variance sensitive to the resolved bandwidth. Whitehouse and Archard [14] demonstrated already in 1970 that the influence of measurement resolution and evaluation procedures on spectral quantities is substantial, a conclusion reinforced systematically in Whitehouse's comprehensive treatment of scale-dependent parameter uncertainty [44]. The practical consequence is that gradient-derived parameters such as $R_{\mathrm{dqx}}$ and $S_{\mathrm{dq}}$ are inherently less robust than amplitude parameters such as $S_q$: their uncertainty scales with the bandwidth of the resolved spectrum rather than with its integral.

The recent Surface-Topography Challenge [45] has provided the most comprehensive empirical evidence for this problem. In this international benchmark study, 153 researchers from 64 institutions characterised two statistically equivalent surfaces using 2 088 independent measurements across a wide range of instruments and techniques. The results confirm that single-scale roughness parameters show wide disagreement across measurement methods, whereas consensus emerges only through scale-dependent descriptors that respect the resolution limits of each technique. This finding directly affects the present framework: since $\langle E \rangle$ is dominated by the gradient variance, any systematic bias in $R_{\mathrm{dqx}}$, whether from the S-filter cutoff $\lambda_s$, the finite-difference stencil, or the instrument transfer function, propagates into $\bar{\beta}$ and thus into $T_{\mathrm{topo}}$.

A rigorous uncertainty budget for the phase-space state variables therefore requires not only the statistical repeatability of the measurement but also a sensitivity analysis with respect to $\lambda_s$, the gradient estimator, and the instrument transfer function. The extension of the framework to time-resolved running-in trajectories with intermediate surface states, and the systematic quantification of how filter bandwidth affects $T_{\mathrm{topo}}$, are the subject of ongoing research.

**Data availability statement:** The data that support the findings of this study are available upon reasonable request from the authors.

**Acknowledgements:** The authors thank Rainer Brodmann for the preliminary work on the angular distribution, Arhia Fatemi (Robert Bosch GmbH) for providing the experimental data, Dorothee Hüser and Poul Erik Hansen for the courage to follow an unconventional line of reasoning, and Andreas Storz for drawing attention to the Surface Topography Challenge.